# Broadband Magnetoresistance in Ferromagnetic and Paramagnetic Samples of $La_{0.7}Ca_{0.3-x}Sr_xMnO_3$

U. Chaudhuri, A. Chanda and R. Mahendiran[1]

Physics Department, 2 Science Drive 3, National University of Singapore,

Singapore-117551, Republic of Singapore

**Abstract**

We have studied the room-temperature magnetoimpedance of paramagnetic ($x = 0.06$) and ferromagnetic ($x = 0.1$) samples in $La_{0.7}Ca_{0.3-x}Sr_xMnO_3$ series using a radio-frequency impedance analyzer and also microwave power absorption using a network analyzer. In both measurements, samples were enclosed tightly inside a copper strip and impedance or reflection coefficient of this copper 'strip coil' was measured as a function of applied magnetic field for different frequencies of current ($f = 0.1$ to 2.5 GHz). The direction of the applied magnetic field was perpendicular to the alternating magnetic field produced by the coil. In the ferromagnetic sample ($x = 0.1$), magnetoresistance shows a peak around zero field for lower frequencies but a peak appears at H > 0 at higher frequencies. The position of the peak shifts towards higher fields with increasing frequency. A similar trend is also found for the paramagnetic sample ($x = 0.06$) but the peak occurs at a higher field compared to the ferromagnetic sample for the same frequency. Microwave power absorption also shows features similar to magnetoresistance. Line shape analysis of the data was performed by fitting the data to a Lorentzian function. It is concluded that the observed features are imprints of ferromagnetic resonance in $x = 0.1$ and paramagnetic resonance in $x = 0.06$ samples.

---

[1] Author for correspondence(phyrm@nus.edu.sg)

## 1. Introduction

The most remarkable property of manganese perovskite oxides ($R_{1-x}A_xMnO_3$ where R = La, Pr, Nd etc., and A = Sr, Ba, Ca) is the occurrence of colossal negative magnetoresistance around ferromagnetic Curie temperature and below. Magnetoresistance is usually measured by a four-probe method with a direct current (*dc*) or a low frequency ($f$ < 500 Hz) current using a lock-in amplifier. On the other hand, magnetoresistance in GHz range is estimated from the changes in microwave reflectivity of a cavity resonator loaded with a manganite sample while an externally applied *dc* magnetic field is swept. The change in microwave reflectivity is related to an alteration in surface resistance of the sample. Dominguez *et al.*[1] and Srinivas *et al.*[2] reported large microwave magnetoresistance at low fields (~80% for H = 600 Oe) compared to magnetoresistance measured with a *dc* current in ferromagnetic samples of $Nd_{0.7}Sr_{0.3}MnO_3$ and $La_{0.7}A_{0.3}MnO_3$ (A= Sr, Ba), respectively. However, those measurements were restricted to a single microwave frequency ($f$ = 9.8 GHz), and multiple cavities had to be used for different frequencies. Very recently, we have reported that the magnitude of magnetoresistance in samples of $La_{0.7}Ca_{0.3-x}Sr_xMnO_3$ series is not only enhanced if a radio frequency (RF)/ microwave (MW) current ($f$ = 0.01 to 3 GHz) is directly passed through the sample, but the field-dependence of high frequency magnetoimpedance ($f$ = 0.1 to 3 GHz) also shows features of ferromagnetic and paramagnetic resonances (FMR and EPR).[3] The possibility of FMR detection via magnetoimpedance was proven for Co-based amorphous alloys but not for EPR[4,5]. Electrical detection of EPR and FMR using a simple magnetoimpedance method that does not make use of microwave cavities or coplanar waveguides is advantageous for applications because of the simplicity of the technique. However, it is important to confirm these results by alternative methods. For this purpose, in this work, we have used two different techniques making use of two different instruments (An RF impedance analyzer and a vector network analyzer).

## 2. Experimental details

Polycrystalline samples of $La_{0.7}Ca_{0.3-x}Sr_xMnO_3$ ($x$ = 0 to 0.3) were prepared by solid state reaction method and characterized earlier by X-ray diffraction, magnetization and magnetoimpedance. Previously, we reported that the ferromagnetic Curie temperature ($T_C$) increased with Sr content[3]. The samples with $x$ > 0.06 are ferromagnetic and $x \leq 0.06$ are paramagnetic at room temperature. We choose a sample of $La_{0.7}Ca_{0.24}Sr_{0.06}MnO_3$ which is paramagnetic at room temperature ($T_C$ = 292 K) and a sample of $La_{0.7}Ca_{0.2}Sr_{0.1}MnO_3$ which is ferromagnetic ($T_C$ =309 K) at room temperature (T=298 K) for the present work. These samples were cut into a rectangular shape (5 mm length x 3 mm width x 2 mm thick). A copper strip of similar dimensions made from 0.2 mm thick copper sheet was folded into a cuboidal coil. The sample was firmly held inside the strip coil. The inner surface of the strip coil was covered with a Kapton tape layer to electrically isolate the sample from the copper strip. The two ends of the copper strips were connected to a radio-frequency impedance analyzer (Agilent model E4991A) or to a vector network analyzer (Agilent model N5230A) using high frequency cables from HUBER+SUHNER. These instruments were calibrated by performing the standard open-short-load procedure. The strip coil with the sample was placed at the center of poles of an electromagnet. Resistance of the strip coil was measured by the impedance analyzer for each *dc* magnetic fields for different frequencies of alternating current (*ac*). Magnetoresistance is calculated from the standard definition: $MR_{ac}$ = [$R(H, f)$-$R(0,f)$]/$R(0,f)$ where $R(H, f)$ is the resistance in a field $H$ for a frequency ($f$) of the current. Data were taken without and with the sample inside the strip coil for same set of frequencies and magnetic fields. Later, the empty coil data was subtracted from the $MR_{ac}$ data taken with the sample. The magnetic field dependence of the microwave power absorption (MWPA) was measured using a vector network analyser (VNA) over a broad frequency range (100 MHz - 2.5 GHz). Here also, the empty coil data were subtracted. Previously, Korenivski et al.[6] had employed a similar copper strip to measure impedance up to 1 GHz using impedance analyzer HP4191A. For MWPA measurement, reflection coefficient of electromagnetic waves in the copper strip is measured via the $S_{11}$ scattering parameter. $S_{11}$ represents the power that is reflected from the device connected to the VNA. To estimate the power absorbed by the sample in the strip coil due to the application of a magnetic field, we record $\Delta P(H) = S_{11}(H, f) - S_{11}(H_{max}, f)$ where, $H_{max}$ is the maximum value of the applied *dc* magnetic field.

## 3. Results

Figs. 1(a) and (b) display the magnetic field dependence of the alternating current (*ac*) magnetoresistance ($MR_{ac}$) for the samples $x = 0.06$ and 0.1, respectively at room temperature in the frequency range: $f = 0.1$ to 2.5 GHz. The $MR_{ac}$ at $f = 0.1$ GHz and 0.5 GHz for the paramagnetic sample $x = 0.06$ displays a single peak at the origin $H_{dc} = 0$. Surprisingly, $MR_{ac}$ for $f > 0.5$ GHz abruptly increases and exhibit a peak at $H_{dc} = \pm H_c$ on either sides of the zero field. Both the peaks at $\pm H_c$ shift towards higher value of $H_{dc}$ with increase in the frequency of the *ac* current. On the other hand, $MR_{ac}$ at $f = 0.1$ GHz for the ferromagnetic sample $x = 0.1$ also shows a single peak at $H_{dc} = 0$. However, as the frequency of the current increases above $f = 1$ GHz, the single peak at $H_{dc} = 0$ splits into two symmetrical peaks at $H_{dc} = \pm H_p$ on either sides of $H_{dc} = 0$. In addition, both these peaks also move towards higher value of $H_{dc}$ as the frequency of the current increases further, similar to the response observed in the paramagnetic sample. However, the peaks in $MR_{ac}$ occurs at higher values of $H_{dc}$ for the paramagnetic sample (0.784 kOe at 2 GHz) in comparison to the ferromagnetic sample ($H_p = 0.402$ kOe at 2 GHz).

Figs. 2 (a) and (b) compare the magnetic field dependence of the normalized power absorption ($\Delta P$) for $x = 0.06$ and 0.1 samples, respectively at room temperature for selected frequencies of the microwave electromagnetic field between $f = 0.1$ GHz and 2.5 GHz. $\Delta P$ at $f = 0.1$ GHz for $x = 0.06$ sample exhibits a single peak at $H_{dc} = 0$ which splits into two symmetrical sharp peaks at $H_{dc} = \pm H_c$ for $f \geq 0.5$ GHz. Both the peaks in $\Delta P$ shift towards higher $H_{dc}$ rapidly as the frequency of the microwave electromagnetic field increases similar to the behavior of $MR_{ac}$. Conversely, $\Delta P$ at $f = 0.1$ GHz for the ferromagnetic sample $x = 0.1$ shows a single peak at $H_{dc} = 0$ but splits into double peaks at $H_{dc} = \pm H_p$ for $f \geq 1.5$ GHz. Similar to the behavior of $MR_{ac}$, both the peaks in $\Delta P$ also move apart from each other towards higher values of $H_{dc}$ as the frequency of the microwave electromagnetic field increases. The peaks in $\Delta P$ are positioned at higher values of $H_{dc}$ for $x = 0.06$ in contrast to $x = 0.1$. Thus, both *ac MR* and $\Delta P$ exhibit distinct features for the ferromagnetic and paramagnetic samples.

## 4. Discussion

The flow of RF current in the copper strip coil generates axial RF magnetic field at the center of the coil and hence the sample's magnetization oscillates with the frequency of the RF magnetic field, which in turns affects the complex impedance ($Z$) of the strip coil, The real ($R$) and imaginary ($X$) components of the complex electrical impedance ($Z$) of the copper strip coil depend on the relative permeability ($\mu_r$) of the sample through the relations $R = K\sqrt{(\omega\mu_0\mu_r'')}$ and $X = K\sqrt{(\omega\mu_0\mu_r')}$, where, $\omega$ is the angular frequency, $\mu_o$ is the free space permeability, $\mu_r'$ and $\mu_r''$ are the real and imaginary components of relative permeability of the sample and $K$ is a constant related to the geometry of the strip coil. Thus, changes in permeability of the sample in response to variations of magnetic field and frequency will give rise to changes in impedance. The RF magnetic field and the applied dc magnetic field ($H_{dc}$) are perpendicular to each other in our experiment as like the field configuration in a conventional microwave cavity based EPR/FMR spectrometer. Both core spins ($t_{2g}^3$: S = 3/2) and $e_g$ electron's spin (S = 1/2) of Mn ions contribute to the total magnetization of the sample. The magnetization re-orients itself to align along the direction of the *dc* magnetic field but the RF magnetic field, which is transverse to the *dc* magnetic field induces precession of the magnetization about the *dc* magnetic field. When resonance condition is met, the sample absorbs a maximum power from the RF electromagnetic field. The power absorbed (*P*) from the RF magnetic field by a sample of volume *V* is $P = \frac{1}{2}V\mu_r^{''}\omega h_{rf}^2$ where, $h_{rf}$ is the amplitude of the RF magnetic field. Magnetic field dependence of $\mu_r^{''}$ goes through a peak value at the resonance field ($H_{res}$), hence, power absorption $P(H_{dc})$ will also show a peak at $H_{res}$. As *R* depends on $\mu_r^{''}$, the anomalous features in $MR_{ac}$ in the ferromagnetic (*x* = 0.1) and paramagnetic (*x* = 0.06) samples can be considered as manifestations of maximum power absorption by the sample due to ferromagnetic resonance (FMR) and electron paramagnetic resonance (EPR), respectively.

Since both absorption and dispersive signals are mixed in power absorption for conducting samples during resonance, we choose to fit the $MR_{ac}$ line shapes for the paramagnetic sample *x* = 0.06 using the following equation,

$$MR_{ac}(H_{dc}) = A_{Sym}\frac{\left(\frac{\Delta H}{2}\right)^2}{(H_{dc}-H_{res})^2+\left(\frac{\Delta H}{2}\right)^2} + A_{Asym}\frac{\frac{\Delta H}{2}(H_{dc}-H_{res})}{(H_{dc}-H_{res})^2+\left(\frac{\Delta H}{2}\right)^2} + A_0 \quad (1)$$

where, $A_{Sym}$ and $A_{Asym}$ are the coefficients of symmetric and antisymmetric Lorentzian functions, $H_{res}$ and $\Delta H$ are the resonance field and linewidth (Full width at the half maximum), respectively and $A_0$ is a constant. The symmetric Lorentzian function accounts for the absorption component and the antisymmetric Lorentzian function is related to the dispersive component[7]. The fitting of the $MR_{ac}$ line shape at $f = 2$ GHz for the $x = 0.06$ sample is shown on the left y-scale of Fig. 3(a). We have also analyzed the normalized microwave power absorption ($\Delta P$ ($H_{dc}$)) line shape for the paramagnetic $x = 0.06$ sample using a linear combination of the symmetric and the antisymmetric Lorentzian functions which can be expressed as,

$$\Delta P(H_{dc}) = P_{Sym} \frac{\left(\frac{\Delta H}{2}\right)^2}{(H_{dc}-H_{res})^2+\left(\frac{\Delta H}{2}\right)^2} + P_{Asym} \frac{\frac{\Delta H}{2}(H_{dc}-H_{res})}{(H_{dc}-H_{res})^2+\left(\frac{\Delta H}{2}\right)^2} + P_0 \quad (3)$$

where, $P_{Sym}$ and $P_{Asym}$ are the coefficients of symmetric and antisymmetric Lorentzian functions and $P_0$ is a constant. The fitting of the $\Delta P$ line shape at $f = 2$ GHz is shown on the right y-scale of Fig. 3(a). $H_{res}$ and $\Delta H$ were obtained using these fits and analyzed. $\Delta H$ for the ferromagnetic $x=0.1$ sample was too broad ($\Delta H > H_{res}$) and could not be obtained using Eq. 2 or Eq. 3. $\Delta H$ for the paramagnetic $x=0.06$ sample was obtained and presented in Fig. 3(b). $\Delta H$ obtained from $MR_{ac}$ measurements (open circles) and $\Delta P$ measurements (closed spheres) both increased with increase in frequency of the current in the strip coil. Solid lines represent the linear relationship of $f$ vs $\Delta H$ which agrees with the Landau-Lifshitz-Gilbert equation for magnetization precession given by Eq. 4.

$$\frac{d\boldsymbol{M}}{dt} = \gamma \mu_0 (\boldsymbol{M} \times \boldsymbol{H}) + \frac{\alpha}{M}\left(\boldsymbol{M} \times \frac{d\boldsymbol{M}}{dt}\right) \quad (4)$$

It can be observed, for the paramagnetic sample, that $H_{res}$ increases linearly with $f$ and hence follows the resonance condition for EPR, i.e., $f_{res} = (\gamma/2\pi) H_{dc}$, where $\gamma$ is the gyromagnetic ratio. A linear fit to the $f_{res}$ vs. $H_{dc}$ curve yields $\gamma/2\pi$ values to be around 2.68±0.05 MHz/Oe using the impedance analyzer method and 2.75±0.02 MHz/Oe using the VNA. These values slightly deviate from the free electron value of 2.8 MHz/Oe. This difference could be due to limitation of the maximum frequency (2.5 GHz) in our experiment compared to commercial EPR spectrometers which usually operate at 9.8 GHz. Depending on measurement parameters, $\gamma/2\pi$ values can be different and a more accurate determination of $\gamma/2\pi$ requires measurements carried out several tens of GHz[8]. Using these $\gamma$ values, $\alpha$ was estimated to be 0.0402 and 0.0550 from the $MR_{ac}$ and $\Delta P$ measurements respectively.

Fig. 3 (c) illustrates the $f_{res}$ vs. $H_{dc}$ curves for the ferromagnetic sample $x = 0.1$. We fitted the $f_{res}$ vs. $H_{dc}$ curves with the Kittel's equation, $f_{res} = \frac{\gamma}{2\pi}\sqrt{[(H_{dc} + H_k)(H_{dc} + H_k + M_S)]}$ where, $H_k$ is the anisotropy field, $M_S$ is the saturation magnetization and $\gamma$ is the gyromagnetic ratio. Fitting the $f_{res}$ vs $H_{dc}$ curves with the Kittel's equation yields $\gamma/2\pi = 2.52\pm0.02$ MHz/Oe for the $MR_{ac}$ line shape and $2.59\pm0.02$ MHz/Oe for the $\Delta P$ line shape. These distinct results for paramagnetic and ferromagnetic samples presented above using the strip coil method where the chosen sample was subjected to RF magnetic field induced by the alternating current in the strip coil indeed confirms our earlier results obtained when RF currents were directly passed through the samples[3]. An added advantage of the strip coil technique is that it can also be used for insulating magnetic samples.

## 5. Summary

In summary, we have investigated the change in electrical impedance in a paramagnetic ($x = 0.06$) and ferromagnetic ($x = 0.1$) samples in $La_{0.7}Ca_{0.3-x}Sr_xMnO_3$ series as a function of $dc$ magnetic field and frequency using a radio-frequency impedance analyzer by measuring the impedance of a strip coil enclosing the sample inside it at room temperature. Our magnetoimpedance results show features of electron paramagnetic resonance and ferromagnetic resonance in $x = 0.06$ and $0.1$, respectively which were validated by microwave absorption measurements using a vector network analyzer. Hence, this low-cost magnetoimpedance technique can potentially investigate spin dynamics in various nanostructured magnetic materials or insulators.

**Acknowledgements:** R. M. thanks the Ministry of Education for supporting this work (Grant number: R144-000-381-112).

# Figure Captions

**FIG. 1** Magnetic field dependence of the *ac* magnetoresistance ($MR_{ac}$) for different frequencies between $f = 0.1$ GHz and 2.5 GHz for the (a) paramagnetic ($x = 0.06$) and (b) ferromagnetic ($x = 0.1$) samples in $La_{0.7}Ca_{0.3-x}Sr_xMnO_3$ series room temperature.

**FIG. 2** Magnetic field dependence of the normalized power absorption ($\Delta P$) for (a) $x = 0.06$ and (b) $x = 0.1$ at room temperature for different frequencies between $f = 0.1$ GHz and 2.5 GHz.

**FIG. 3** (a) Lorentzian fits to the line shape of $MR_{ac}$ (left y-scale) and $\Delta P$ (right y-scale) for the $x = 0.06$ sample at $f = 2$ GHz, (b) Line width ($\Delta H$) as a function of frequency ($f$), obtained from the line shape analysis of $MR_{ac}$ (open circles) and $\Delta P$ (closed spheres) measurements. The solid lines are the linear fits to the data. (c) Plot of $f_{res}$ vs $H_{dc}$ obtained from the $MR_{ac}$ (open symbol) and $\Delta P$ (closed symbol) line shape analysis for $x = 0.06$ (circles) and 0.1 (squares) samples. The solid lines depict the Kittel equation fitting for all the samples.

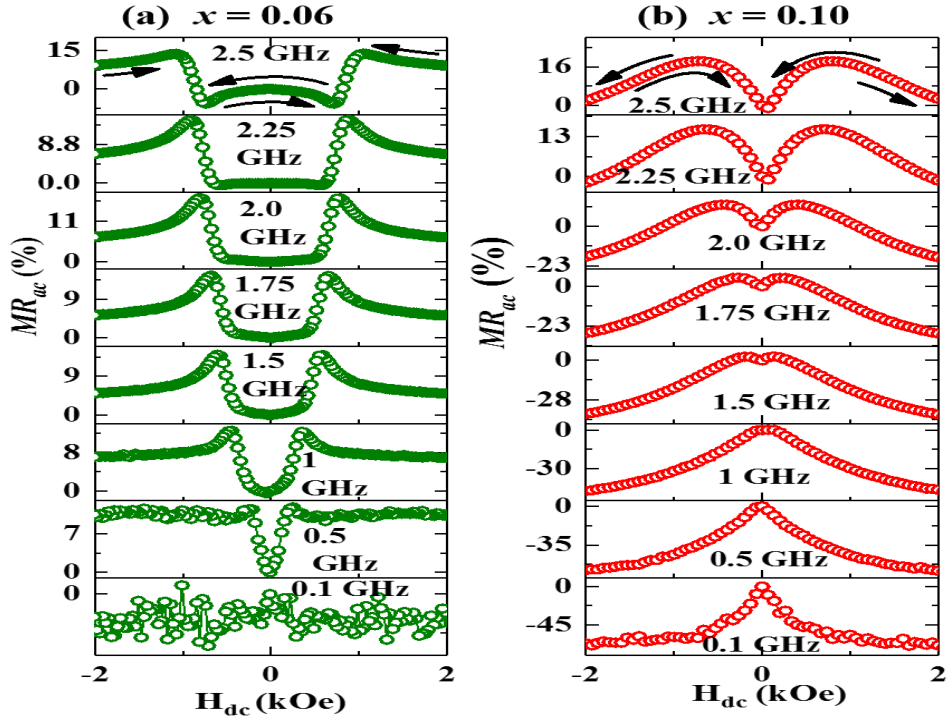

**FIG. 1** Magnetic field dependence of the *ac* magnetoresistance ($MR_{ac}$) for different frequencies between $f = 0.1$ GHz and 2.5 GHz for the (a) paramagnetic ($x = 0.06$) and (b) ferromagnetic ($x = 0.1$) samples in $La_{0.7}Ca_{0.3-x}Sr_xMnO_3$ series room temperature.

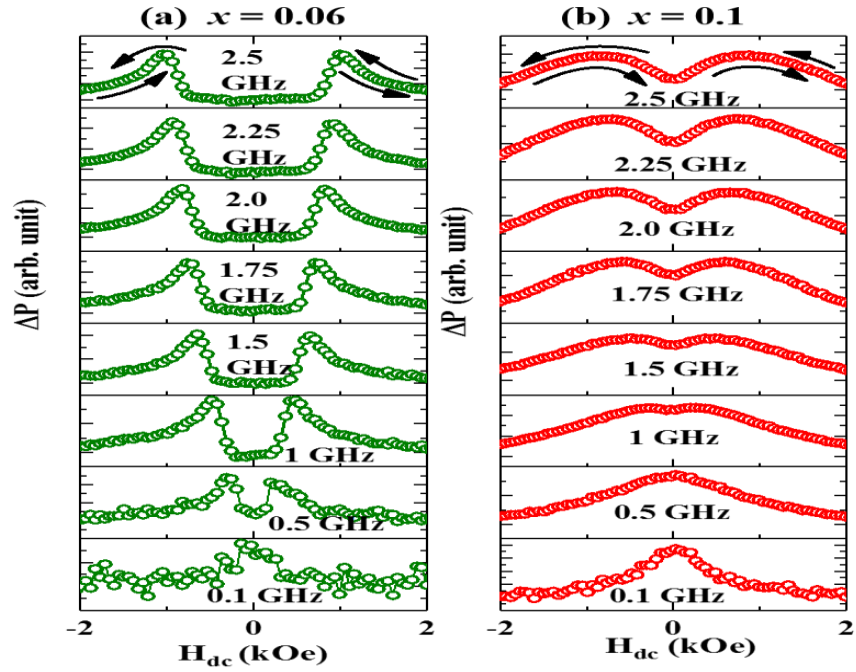

**FIG. 2** Magnetic field dependence of the normalized power absorption ($\Delta P$) for (a) $x = 0.06$ and (b) $x = 0.1$ at room temperature for different frequencies between $f = 0.1$ GHz and 2.5 GHz.

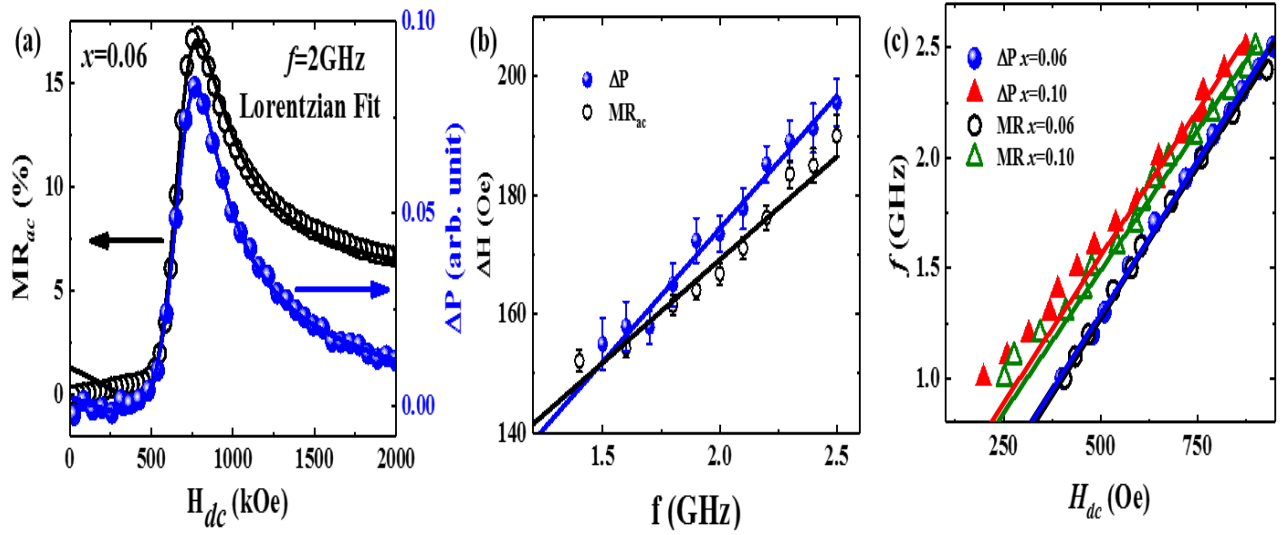

**FIG. 3** (a) Lorentzian fits to the line shape of $MR_{ac}$ (left y-scale) and $\Delta P$ (right y-scale) for the $x = 0.06$ sample at $f = 2$ GHz, (b) Line width ($\Delta H$) as a function of frequency ($f$), obtained from the line shape analysis of $MR_{ac}$ (open circles) and $\Delta P$ (closed spheres) measurements. The solid lines are the linear fits to the data. (c) Plot of $f_{res}$ vs $H_{dc}$ obtained from the $MR_{ac}$ (open symbol) and $\Delta P$ (closed symbol) line shape analysis for $x = 0.06$ (circles) and 0.1 (squares) samples. The solid lines depict the Kittel equation fitting for all the samples.